\documentclass[twocolumn,aps,pra,showpacs,groupedaddress]{revtex4}
\usepackage{epsfig,amssymb,amsmath,mathrsfs,color}
\usepackage[active]{srcltx}
\usepackage[matrix,frame,arrow]{xypic}
\usepackage[matrix,frame,arrow]{xy}
\usepackage{amsmath}
\newcommand{\bra}[1]{\left\langle{#1}\right\vert}
\newcommand{\ket}[1]{\left\vert{#1}\right\rangle}
\newcommand{\qw}[1][-1]{\ar @{-} [0,#1]}

\newcommand{\cw}[1][-1]{\ar @{=} [0,#1]}

\newcommand{\gate}[1]{*{\xy *+<.6em>{#1};p\save+LU;+RU **\dir{-}\restore\save+RU;+RD **\dir{-}\restore\save+RD;+LD **\dir{-}\restore\POS+LD;+LU **\dir{-}\endxy} \qw}

\newcommand{\measureD}[1]{*{\xy*+=+<.5em>{\vphantom{\rule{0em}{.1em}#1}}*\cir{r_l};p\save*!R{#1} \restore\save+UC;+UC-<.5em,0em>*!R{\hphantom{#1}}+L **\dir{-} \restore\save+DC;+DC-<.5em,0em>*!R{\hphantom{#1}}+L **\dir{-} \restore\POS+UC-<.5em,0em>*!R{\hphantom{#1}}+L;+DC-<.5em,0em>*!R{\hphantom{#1}}+L **\dir{-} \endxy} \qw}

\newcommand{\multimeasureD}[2]{*+<1em,.9em>{\hphantom{#2}}\save[0,0].[#1,0];p\save !C *{#2},p+LU+<0em,0em>;+RU+<-.8em,0em> **\dir{-}\restore\save +LD;+LU **\dir{-}\restore\save +LD;+RD-<.8em,0em> **\dir{-} \restore\save +RD+<0em,.8em>;+RU-<0em,.8em> **\dir{-} \restore \POS !UR*!UR{\cir<.9em>{r_d}};!DR*!DR{\cir<.9em>{d_l}}\restore \qw}

\newcommand{\multigate}[2]{*+<1em,.9em>{\hphantom{#2}} \qw \POS[0,0].[#1,0];p !C *{#2},p \save+LU;+RU **\dir{-}\restore\save+RU;+RD **\dir{-}\restore\save+RD;+LD **\dir{-}\restore\save+LD;+LU **\dir{-}\restore}
\newcommand{\ghost}[1]{*+<1em,.9em>{\hphantom{#1}} \qw}

\newcommand{\ustick}[1]{*!D!<0em,-.5em>=<0em>{#1}}

\newcommand{\Qcircuit}[1][0em]{\xymatrix @*=<#1>}

\newcommand{\pureghost}[1]{*+<1em,.9em>{\hphantom{#1}}}
\newcommand{\multiprepareC}[2]{*+<1em,.9em>{\hphantom{#2}}\save[0,0].[#1,0];p\save !C
  *{#2},p+RU+<0em,0em>;+LU+<+.8em,0em> **\dir{-}\restore\save +RD;+RU **\dir{-}\restore\save
  +RD;+LD+<.8em,0em> **\dir{-} \restore\save +LD+<0em,.8em>;+LU-<0em,.8em> **\dir{-} \restore \POS
  !UL*!UL{\cir<.9em>{u_r}};!DL*!DL{\cir<.9em>{l_u}}\restore}
\newcommand{\prepareC}[1]{*{\xy*+=+<.5em>{\vphantom{#1\rule{0em}{.1em}}}*\cir{l^r};p\save*!L{#1} \restore\save+UC;+UC+<.5em,0em>*!L{\hphantom{#1}}+R **\dir{-} \restore\save+DC;+DC+<.5em,0em>*!L{\hphantom{#1}}+R **\dir{-} \restore\POS+UC+<.5em,0em>*!L{\hphantom{#1}}+R;+DC+<.5em,0em>*!L{\hphantom{#1}}+R **\dir{-} \endxy}}

\newcommand{\multipuregate}[2]{*+<1em,.9em>{\hphantom{#2}}
\POS[0,0].[#1,0];p !C *{#2},p \save+LU;+RU
**\dir{-}\restore\save+RU;+RD **\dir{-}\restore\save+RD;+LD
**\dir{-}\restore\save+LD;+LU **\dir{-}\restore}

\newcommand{\puremeasureD}[1]{*{\xy*+=+<.5em>{\vphantom{#1}}*\cir{r_l};p\save*!R{#1} \restore\save+UC;+UC-<.5em,0em>*!R{\hphantom{#1}}+L **\dir{-} \restore\save+DC;+DC-<.5em,0em>*!R{\hphantom{#1}}+L **\dir{-} \restore\POS+UC-<.5em,0em>*!R{\hphantom{#1}}+L;+DC-<.5em,0em>*!R{\hphantom{#1}}+L **\dir{-} \endxy}}

 \newtheorem{proposition}{Proposition}
 \newtheorem{theorem}{Theorem}
 \newtheorem{definition}{Definition}

 \def\Lin{{\mathcal L}} 
\def\vec#1{\boldsymbol{#1}} \def\qed{$\blacksquare$}  \def\Proof{\medskip\par\noindent{\bf Proof.  }}
\def\>{\rangle} \def\<{\langle}\def\kk{\>\! \>}\def\bb{\<\!\<}

  \def\Tr{{\rm Tr}}
  
\def\defset#1{\mathsf{#1}} 
\def\rank{\operatorname{Rnk}}   
  
\def\supp{\operatorname{Supp}}

 \newcommand{\Ket}[1]{| #1 \rangle\! \rangle} 
\newcommand{\Bra}[1]{\langle \! \langle #1 |}

\newcommand{\KetBra}[2]{\Ket{#1} \Bra{#2}}

\newcommand{\ketbra}[2]{\ket{#1} \bra{#2}}
\newcommand{\hilb}[1]{\mathcal{#1}}

\begin{document}

\title{Memory cost of quantum protocols}
\author{Alessandro Bisio}\email{alessandro.bisio@unipv.it}
\affiliation{{\em QUIT Group}, Dipartimento di Fisica ``A. Volta'' and
INFN, via Bassi 6, 27100 Pavia, Italy} \homepage{http://www.qubit.it}
\author{Giacomo Mauro D'Ariano}\email{dariano@unipv.it}
\affiliation{{\em QUIT Group}, Dipartimento di Fisica ``A. Volta'' and
INFN, via Bassi 6, 27100 Pavia, Italy} \homepage{http://www.qubit.it}
\author{Paolo Perinotti}\email{paolo.perinotti@unipv.it}
\affiliation{{\em QUIT Group}, Dipartimento di Fisica ``A. Volta'' and
INFN, via Bassi 6, 27100 Pavia, Italy} \homepage{http://www.qubit.it}
\author{Michal Sedl\'ak}\email{michal.sedlak@unipv.it}
\affiliation{{\em QUIT Group}, Dipartimento di Fisica ``A. Volta'',
via Bassi 6, 27100 Pavia, Italy}
 \homepage{http://www.qubit.it}
 \affiliation{Institute of Physics, Slovak Academy of Sciences, D\'ubravsk\'a cesta 9, 845 11 Bratislava, Slovakia}
\date{\today}
\begin{abstract}
  In this paper we consider the problem of minimizing the ancillary
  systems required to realize an arbitrary strategy of a quantum
  protocol, with the assistance of classical memory. For this
  purpose we 
  introduce the notion of {\em memory cost} of a
  strategy, which measures the resources required  in
  terms of ancillary dimension. We provide a condition for the cost to
  be equal to a given value, and we use this result to evaluate the cost
  in some special cases. 
  As an example we show that any covariant protocol for the cloning of
  a unitary transformation requires at most one ancillary qubit.
  We also prove that the memory cost has to be determined globally, and cannot be
  calculated by optimizing the resources independently at each step
  of the strategy.
\end{abstract} \maketitle

\section{Introduction}

Since the advent of Quantum Computation, the most important
theoretical efforts in this field were aimed to prove a computational
speedup in many information processing tasks \cite{Grover,shor} with
respect to the classical counterparts. For this reason, the optimization
of algorithms is typically aimed at minimizing the number of
computational steps, possibly at the expense of the
computational space, i.e. the amount of ancillary quantum systems
(qubits) that are needed in the computation. This
choice is dictated by the fact that time is the most valuable resource
in computation. Moreover,
compared with the classical case,
 in Quantum Computation time optimization is
even more important because of the detrimental effects
of decoherence.

Beside time minimization, next priority is the optimization of the
computational space. More precisely, the resource we need to minimize
is quantum memory, that is the number of ancillary systems that need
to be kept coherent between subsequent steps. 
Since a classical memory has a negligible cost with respect to a
quantum one,
it would be very valuable to replace part of the quantum
memory by a classical channel.

In Ref. \cite{algorithm} the minimization of the memory was carried
out under the restrictive assumption that all the ancillary systems
introduced during the computation are kept coherent until the very
last step. In the present paper, we consider the same problem, taking
into account the possibility of breaking the coherence of ancillary
systems during the computation without affecting the overall
strategy, by measuring and compressing the ancillary computational
space at the expense of an extra classical memory carrying measurement
outcomes. 
In order to measure the quantum  memory
cost of a strategy we introduce the notion of {\em
memory cost}  which will be the logarithm
of the maximum 
dimension of ancillary quantum systems required at all steps.
For the special case of a strategy describing a single channel, our
notion of memory cost coincides with the one of entanglement cost
recently introduced in Ref.  \cite{zurich}. Indeed,
a single channel can be interpreted as a quantum strategy made of
two steps: i)
a quantum instrument 
followed by a compression conditional on the classical outcome at
first step
and ii) a conditional decompression at the second step.
After providing a
necessary and sufficient condition for a strategy to have a given
memory cost, we show that its optimization
 cannot generally be carried out by minimizing the memory required
at each step separately. The reason for this is that in
the memory optimization of a strategy one can exploit different
channel implementation of the same comb. 
This fact implies
that in general the optimization must be a global one. Finally, we
 investigate how the symmetry properties of a quantum strategy
can lead to nontrivial a bound of its memory cost and we calculate
it  for simple classes of covariant channels.

The paper is organized as follows. In Section \ref{sec:prel-notat} we
present some elementary result of linear algebra with special emphasis
on the Choi isomorphism.  In Section \ref{sec:prelim} we review the
general theory of Quantum combs \cite{QCA, comblong, actaphys},
which provides a unified framework to treat quantum strategy.
Section \ref{sec:entcost} provides the definition of memory cost
along with the main theorem. In Section \ref{sec:examples} we provide
some examples in which the application of the necessary and sufficient
condition allows us to draw non-trivial conclusions about the cost of
strategy. We conclude the paper with Section \ref{sec:concl} where we
summarize the results and discuss some open problems.

\section{Preliminaries and notation}\label{sec:prel-notat}

In this section we introduce the  basic mathematical tools and the
notation that will be used throughout the whole manuscript.
If $\hilb{H}$ denotes a finite dimensional Hilbert spaces
$\mathcal{L}(\hilb{H})$ denotes the set of linear operators on
$\hilb{H}$.
Once we fixed an orthonormal basis $\{ \ket{n} \}$
for $\hilb{H}$ the following one to one correspondence between 
$\mathcal{L}(\hilb{H})$ and  $\hilb{H} \otimes \hilb{H} $
is well defined:
\begin{align}
  \label{eq:doubleket}
  &A = \sum_{nm}\bra{n}A\ket{m} \ketbra{n}{m} 
\leftrightarrow
\Ket{A} = \sum_{nm}\bra{n}A\ket{m} \ket{n}\ket{m}& \nonumber\\
&A \otimes B \Ket{C} = \Ket{ACB^T},&
\end{align}
where $A^T$ denotes transposition
of $A$ with respect to the fixed basis
($A^*$ will denote the complex conjugation).
In the following we will denote 
$\supp(A)$ the support of  $A$
and $\rank(A)$ the dimension of 
$\supp(A)$, i.e. $\rank(A) := \dim(\supp(A))$
The set of linear maps from
$\mathcal{L}(\hilb{H}_1)$ to $\mathcal{L}(\hilb{H}_2)$ will be denoted
by $\mathcal{L}(\mathcal{L}(\hilb{H}_1), \mathcal{L}(\hilb{H}_2))$.

The following result, due to Choi \cite{Choi}, 
establishes a bijective correspondence between
$\mathcal{L}(\mathcal{L}(\hilb{H}_1), \mathcal{L}(\hilb{H}_2))$
and  $\mathcal{L}(\hilb{H}_1 \otimes \hilb{H}_2)$.
\begin{theorem}\label{th:choiisomor}
Let $\mathcal{I}$ be the identical map on $\mathcal{L}(\hilb{H}_1)$.
The linear map $\mathfrak{C}:\mathcal{L}(\mathcal{L}(\hilb{H}_1),
\mathcal{L}(\hilb{H}_2)) \to \mathcal{L}(\hilb{H}_1 \otimes
\hilb{H}_2))$ defined as
\begin{align}
  \label{eq:choiisomorphism}
  \mathfrak{C} : \mathcal{C} \mapsto C := \mathcal{C} \otimes
  \mathcal{I} (\KetBra{I}{I}),
\end{align}
is invertible and its inverse
$\mathfrak{C}^{-1}$  
is given by
 \begin{align}
   \label{eq:inversechoi}
   [\mathfrak{C}^{-1}(C)](A) = \Tr_1[ (I_2 \otimes A^T) C] = \mathcal{C}(A),
 \end{align}
where $\Tr_1$ denotes the partial trace over $\hilb{H}_1$ and $I_2$
is the identity operator on $\hilb{H}_2$.
The operator $C = \mathfrak{C}(\mathcal{C}) $
is called \emph{Choi operator} of $\mathcal{C}$.
\end{theorem}
Throughout this paper we will use
the calligraphic style $\mathcal{C}$
to denote the linear map 
and the italic $C$ to denote
the corresponding Choi operator.
It is useful to give a diagrammatic representation of linear maps:
we will sketch a map 
$\mathcal{C} \in
\mathcal{L}
(\mathcal{L}(\bigotimes_{i=1}^N\hilb{H}_{i} ) ,
  \mathcal{L}(\bigotimes_{j=1}^M\hilb{H}_{j}))$ 
as a box with \emph{N input wires} on the left and
\emph{M output wires} on the right, for example if
$\mathcal{C} \in
\mathcal{L}
(\mathcal{L}(\hilb{H}_{0} \otimes \hilb{H}_{0'} ) ,
  \mathcal{L}(\hilb{H}_{1} \otimes \hilb{H}_{1'})$ we have
\begin{align}
  \label{eq:mapasbox}
  C
=
\quad  \begin{aligned}
    \Qcircuit @C=1em @R=1em {
    \ustick{0}&\multigate{1}{C}&\ustick{1}\qw&\\
    \ustick{0'}&\ghost{C}&\ustick{1'}\qw&&&}    
  \end{aligned}
\end{align}
We now 
show
how some features of a linear map $\mathcal{C}$
translates in terms of the Choi operator $C$
\begin{proposition}\label{th:linkproduct}
Let $\mathcal{C} \in \mathcal{L}(\mathcal{L}(\hilb{H}_0) ,
  \mathcal{L}(\hilb{H}_1\otimes \hilb{H}_A))$ 
and
$\mathcal{D} \in \mathcal{L}(\mathcal{L}(\hilb{H}_2 \otimes \hilb{H}_A) ,
  \mathcal{L}(\hilb{H}_3))$
be two linear maps and $C$, $D$
be their Choi operators.
Then we have:
\begin{itemize}
\item $\mathcal{C}$ is completely positive 
if and only if $C \geq  0$;
\item $\mathcal{C}$ is trace non increasing 
if and only if $\Tr_{1A}[C] \leq  I_0$ with equality 
when $\mathcal{C}$ is trace preserving;
\item the Choi operator of the composition 
$(\mathcal{I}_1 \otimes \mathcal{D}) \circ ( \mathcal{I}_2
\otimes\mathcal{C} ) $
is given by the \emph{link product} \cite{QCA} of $C$ and $D$,
that is
$\mathfrak{C} ((\mathcal{I}_2 \otimes \mathcal{D}) \circ ( \mathcal{I}_3
\otimes\mathcal{C} )) =  C * D $ where 
\begin{align}
  \label{eq:linkproduct}
  & C * D :=
  \Tr_A[(C \otimes I_{34})(I_{01} \otimes D )]
\end{align}
\end{itemize}
\end{proposition}
The link $C * D$ in Eq. \eqref{eq:linkproduct} can be visualized
as follows:
\begin{align}
C * D =
\begin{aligned}
\Qcircuit @C=1em @R=1em {
    \ustick{0}&\multigate{1}{C}&\ustick{1}\qw&&\ustick{2}&\multigate{1}{D}&\ustick{3}
    \qw\\
    &\pureghost{C}&\qw&\ustick{A}\qw&\qw&\ghost{D}&&}    
  \end{aligned}.\nonumber
\end{align}

\section{Quantum Strategies}\label{sec:prelim}

In the usual description of Quantum Mechanics each physical system is
associated with a Hilbert space $\hilb{H}$ and the states of the
system are represented by positive semi-definite operators $\rho$ with
$\Tr[\rho]=1$.  A single use \cite{note1} of a physical device which
performs a transformation of the system is represented by a linear map
$\mathcal{C} \in \mathcal{L}(\mathcal{L}(\hilb{H}_{\rm in}) ,
\mathcal{L}(\hilb{H}_{\rm out}) )$ which is completely positive ($C
\geq 0$) and trace non increasing ($\Tr_{\rm out} C \leq I_{\rm in}$).
If the transformation is deterministic $\mathcal{C}$ is trace
preserving ($\Tr_{\rm out}[C]= I_{\rm in}$) and is called
\emph{quantum channel}, while in the general probabilistic case it is
called \emph{quantum operation}.  A set of quantum operations
$\vec{{\mathcal{M}}}\equiv \{ \mathcal{M}_i\}$ such that
$\mathcal{M}_{\Omega} := \sum_i \mathcal{M}_i$ is a quantum channel is
called \emph{quantum instrument}. Physically, a quantum instrument
describes a device that has both a classical and a quantum outcome.
One can regard a demolishing measurement device as a special case of
quantum instrument where there is only a classical outcome.  The
mathematical description of a measurement is given in this case by a
set of positive operators $\vec{P} := \{ P_i \}$ which sum up to the
identity $\sum_i P_i = I$ ---a \emph{positive operator valued measure}
(POVM).

A general \emph{quantum strategy} can be obtained
by connecting the outputs of some transformations into the input of some
others. 
If the transformation that we are connecting are deterministic,
i.e. quantum channels, we have a \emph{deterministic quantum strategies} and
we talk about \emph{probabilistic quantum strategies} otherwise. 
In a valid quantum strategy no closed loops are allowed
\cite{note2}: this requirement
ensures that causality is preserved, since 
a closed path would correspond to a time loop.
Quantum strategies can be used to describe a huge variety
of multi-step quantum protocols, like
cryptographic protocols \cite{bitcommitment, pironio},
standard quantum algorithms \cite{deutschjoz, Grover, shor} and
multi-round quantum games \cite{watrousgame}.

It is possible to prove that any deterministic quantum strategy 
is equivalent to a concatenation  of $N$
channels $\mathcal{C}_i \in \mathcal{L}(\mathcal{L}(\mathcal{H}_{2i-2} \otimes
\mathcal{A}_{i-1}), \mathcal{L}( \mathcal{H}_{2i-1} \otimes \mathcal{A}_{i}))$
($\mathcal{A}_{0} = \mathcal{A}_{N} = \mathbb{C}$)
and thus it is represented
by a map $ \mathcal{R}^{(N)} $ whose Choi operator is given by the link  of the
${C}_i$'s, i.e.
\begin{align}
  \label{eq:quantum Network}
R^{(N)} = C_1 * \dots * C_N.
\end{align}
This result 
allows us to represent
each deterministic  quantum strategy $ \mathcal{R}^{(N)} $
as a sequence of $N$ computational steps
each of them corresponding to a channel
$\mathcal{C}_i$:
\begin{align}
\nonumber\\
 R^{(N)} =   
\begin{aligned}
   \Qcircuit @C=0.9em @R=1.3em {
    \ustick{0}&\multigate{1}{C_1}&\ustick{1}\qw&&\ustick{2}&\multigate{1}{C_2}&\ustick{3}\qw&\pureghost{\dots}&\ustick{2N-2}&\multigate{1}{C_N}&\ustick{2N-1}\qw\\
    &\pureghost{C_1}&\qw& \ustick{\mathcal{A}_1} \qw&\qw&\ghost{C_2}&
\ustick{\mathcal{A}_2} \qw&\cdots&
\ustick{\!\!\!\!\!\!\mathcal{A}_{N-1}}&
\ghost{C_N}&}
  \end{aligned}.\label{Eq:comb}
\end{align}
Eq. \eqref{Eq:comb} is our standard representation of a quantum
strategy $ \mathcal{R}^{(N)} $, where 
the apex $(N)$ makes explicit  the number of steps of the
strategy.
We chose to attach one free incoming and one free outgoing wire to
each map $\mathcal{C}_i$ since strategies in which some input/output
wires are missing corresponds to the special cases in which $\dim(\hilb{H}_j)=1$
for some $j$.
It is worth noticing that a quantum channel $\mathcal{C}$
can be seen either as a single step strategy
$\begin{aligned}
   \Qcircuit @C=0.4em @R=0.6em {
&\gate{C}&\qw&    
}
  \end{aligned}$
or as a two steps strategy
in which both the output of the first
step and the input of the second one are one dimensional:
\begin{align}
  \label{eq:channelasnetw}
  C = 
  \begin{aligned}
    \Qcircuit @C=1.2em @R=1em {
    \ustick{0}&\multigate{1}{C_1}&&\multipuregate{1}{C_2}&\ustick{3}\qw&\\
    &\pureghost{C_1}&\ustick{\mathcal{A}_1}\qw&\ghost{C_2}&&}
  \end{aligned}.
\end{align}
The representation given by Eq. \eqref{eq:channelasnetw}
will be useful when
 discussing the memory cost of a channel.
In Eq. \eqref{Eq:comb} we also chose to label
the free input/output wires by integer numbers.
In this way the 
Hilbert spaces of the input wires are labeled by even numbers
while the output ones correspond to odd numbers.
We define the overall input space of a quantum strategy
$\mathcal{R}^{(N)}$
 as $\hilb{H}_{{\rm in}} = \bigotimes_{i=1}^{N} \hilb{H}_{2i-2}$
and the overall output space as
$\hilb{H}_{{\rm out}} = \bigotimes_{j=1}^N \hilb{H}_{2i-1}$.

The previous considerations
can be summarized by the following definition. 
\begin{definition}\label{def:quantumnetwork}
A linear map $\mathcal{R}^{(N)} \in \mathcal{L}
(
\mathcal{L}(\hilb{H}_{\rm in}), 
\mathcal{L}(\hilb{H}_{\rm out})$
is a deterministic quantum strategy when there exists
a set of channels
$\{ \mathcal{C}_i \in
\mathcal{L}(\mathcal{L}(\mathcal{H}_{2i-2} \otimes
\mathcal{A}_{i-1}), \mathcal{L}( \mathcal{H}_{2i-1} \otimes \mathcal{A}_{i}))\}$  such that
$    C_1 * \dots * C_N = R^{(N)}$.
The set $\boldsymbol{\mathscr{C}} := \{ \mathcal{C}_i\}$ is called a \emph{realization} of
$\mathcal{R}^{(N)}$
and the set  $\mathsf{S} := \{ 1, 2, \dots, N \}$ is
called the \emph{set of steps}
of the quantum strategy.
\end{definition}
It is important to notice that the same ${R}^{(N)}$
can have different realizations.
As far as one is not interested in the 
inner structure of a quantum strategy
but just in its properties
 as a linear map from
$\hilb{H}_{\rm in}$
 to
$\hilb{H}_{\rm out}$,
 the description provided by ${R}^{(N)}$ is exhaustive
and there is no need to specify a realization.
On the other hand, if we fix a realization $\boldsymbol{\mathscr{C}}$ of $\mathcal{R}^{(N)}$
we specify some details about the physical implementation 
of the quantum strategy.
For example, the dimensions of the spaces $\mathcal{A}_i$
determine
the amount of memory
 used in the physical implementation  
of the strategy.

Definition \ref{def:quantumnetwork} identifies the set of  
the Choi operators of deterministic quantum strategies
with the set of linear maps $\mathcal{R}^{(N)}$
for which there exists a realization $\boldsymbol{\mathscr{C}}$.
The following theorem recasts this characterization
 in terms of linear constraints
which $R^{(N)}$ has to  fulfill.
\begin{theorem}\label{th:cornerstonetheorem}
A positive operator 
$R^{(N)}  \in \Lin (\hilb{H}_{\rm in} \otimes \hilb{H}_{\rm out})$
is the Choi operator of a deterministic quantum strategy if and only if
it satisfies the normalization
\begin{equation}\label{recnorm}
\Tr_{2k-1} [ R^{(k)}] = I_{2k-2} \otimes R^{(k-1)} \qquad k=1, \dots, N~
\end{equation}
where $R^{(k)}
\in\Lin (\bigotimes_{n=0}^{2k-1} \hilb{H}_n)$ is the
Choi operator of the reduced quantum strategy with $k$ steps
and $R^{(0)} = 1$.
The Choi operator of a deterministic quantum strategy is called
\emph{deterministic quantum comb}
\cite{QCA}.
\end{theorem}
Theorem \ref{th:cornerstonetheorem} can be understood as
a generalization of Theorem 
\ref{th:choiisomor} to quantum strategies. It provides a
a one to one correspondence between the set of deterministic 
quantum strategies
 and the set of positive semi-definite operators satisfying
Eq. \eqref{recnorm}.

We now extend the previous discussion  
to the probabilistic case.
It is possible to prove a probabilistic counterpart
of Theorem \ref{th:cornerstonetheorem}
which
states that a linear map $\mathcal{S}^{(N)}$
is a probabilistic quantum strategy
if and only if its Choi operator 
${S}^{(N)}$
satisfies the following constraint:
\begin{align}
  \label{eq:probcomb}
  0 \leq S^{(N)} \leq R^{(N)}
\end{align}
where $R^{(N)}$ is a deterministic comb.
The Choi operator of a probabilistic quantum strategy
is called probabilistic quantum comb.
The quantum strategy generalization of
a quantum instrument is called \emph{generalized instrument} and it is
a set of probabilistic quantum strategy
$\vec {\mathcal{R}^{(N)}} := \{ \mathcal{R}^{(N)}_i\}$
such that the set $\vec{R^{(N)}} := \{ R^{(N)}_i \}$
of the corresponding probabilistic quantum  combs satisfies
\begin{align}
  \label{geninst}
  \sum_i R^{(N)}_i = R^{(N)}_\Omega
\end{align}
where $R^{(N)}_\Omega$ is a deterministic quantum comb.
A generalized instrument is the
mathematical representation of a strategy
 that produces both the classical outcome $i$
and the quantum outcome $\mathcal{R}^{(N)}_i(\rho) \in \mathcal{L}(\hilb{H}_{out})$
with probability $\Tr[\mathcal{R}^{(N)}_i(\rho)]$ when the state $\rho
\in \mathcal{L}(\hilb{H}_{in})$ is fed into the free inputs of the strategy.
A typical example of a generalized instrument is a Quantum Network in which at least
one of the devices is a quantum instrument:
\begin{align} \label{exgeninst}
\begin{aligned}
    \Qcircuit @C=1em @R=1em {
      &\ustick{0}&\multigate{1}{C} &\ustick{1} \qw&&
      \ustick{2}&\ghost{\vec{D}} &\ustick{3} \qw
      &&\ustick{4}&\multigate{1}{E} &\ustick{5} \qw&\\
      &&\pureghost{C} &\qw& \ustick{\mathcal{A}_1} \qw & \qw
      &\multigate{-1}{\vec{D}}& \qw& \ustick{\mathcal{A}_2} \qw &\qw&\ghost{E} &}
  \end{aligned}.
\end{align}
In Eq. \eqref{exgeninst} the two channels $C$ and $E$
are connected through wires $\mathcal{A}_1$ and $\mathcal{A}_2$ to the
quantum instrument $\vec{\mathcal{D}}$.
If $\vec {\mathcal{R}^{(N)}} := \{ \mathcal{R}^{(N)}_i\}$
is generalized instrument, one can verify that
$\sum_i {R}^{(N)}_i \otimes \ketbra{i}{i}_E$,
where $\{ \ket{i} _E\}$ is an orthonormal basis
for an ancillary Hilbert space $\mathcal{E}$,
is a deterministic comb.
If we apply the von Neumann measurement
$\vec{P}:= \{ \ketbra{i}{i} \}$ on the ancilla $\mathcal{E}$,
depending on the outcome $i$ the Choi
operator of the strategy will be ${R}^{(N)}_i$.
This proves that every generalized instrument can be realized
as a deterministic quantum strategy followed by a POVM on
an ancillary Hilbert space, i.e.
\begin{align}
\begin{aligned}
   \Qcircuit @C=0.9em @R=1.3em {
    \ustick{0}&\multigate{1}{C_1}&\ustick{1}\qw&&\ustick{2}&\multigate{1}{C_2}&\ustick{3}\qw&\pureghost{\dots}&\ustick{2N-2}&\multigate{1}{C_N}&\ustick{2N-1}\qw\\
    &\pureghost{C_1}&\qw& \ustick{\mathcal{A}_1} \qw&\qw&\ghost{C_2}&
\ustick{\mathcal{A}_2} \qw&\cdots&
\ustick{\!\!\!\!\!\!\mathcal{A}_{N-1}}&
\ghost{C_N}&\ustick{\mathcal{E}}\qw&\measureD{\vec{P}}}
  \end{aligned}.\label{Eq:geninst}
\end{align}
A  generalized instrument such that
$\dim (\mathcal{H}_0) = \dim (\mathcal{H}_{2N+1}) = 1$
is called \emph{tester} and can be interpreted as the quantum strategy
analog
of a POVM.
Specializing Eq. \eqref{Eq:geninst}, we have that
a tester can be realized as a quantum strategy
whose first step is a state preparation 
and the last step is a POVM:
\begin{align}
\begin{aligned}
   \Qcircuit @C=0.9em @R=1.3em {
    &\multiprepareC{1}{\rho}&\ustick{1}\qw&&\ustick{2}&\multigate{1}{C_2}&\ustick{3}\qw&\pureghost{\dots}&\ustick{2N-2}&\multimeasureD{1}{\vec{P}}&\\
    &\pureghost{\rho}&\qw& \ustick{\mathcal{A}_1} \qw&\qw&\ghost{C_2}&
\ustick{\mathcal{A}_2} \qw&\cdots&
\ustick{\!\!\!\!\!\!\mathcal{A}_{N-1}}&
\ghost{\vec{P}}&}
  \end{aligned}.\label{Eq:tester}
\end{align}

Since a quantum strategy is a map from multiple input spaces to
multiple output spaces, we can imagine to
connect two quantum strategies
$\mathcal{R}^{(N)}$ and $\mathcal{S}^{(M)}$
by linking some
outputs of  $\mathcal{R}^{(N)}$ ($\mathcal{S}^{(M)}$) with some inputs of 
$\mathcal{S}^{(M)}$ ($\mathcal{R}^{(N)}$), for example
\begin{align}
R ^{(2)} * S ^{(2)} =  \begin{array}{c}
\!\!\!\! R ^{(2)} \\
\!\!\!\! \overbrace{ \quad \qquad \qquad \qquad }
\\   
\begin{aligned}
     \Qcircuit @C=1em @R=0.7em {
&& &\pureghost{ {C}'_1}  
&\qw &\multigate{1}{ {C}'_2}  &&&&\\
&\multigate{1}{ {C}_1}& \qw  &\multigate{-1}{ {C}'_1}   
&&
\pureghost{ {C}'_2}   & \qw &\multigate{1}{ {C}_2}&\qw\\
& \pureghost{ {C}_1}&   \qw  & 
\qw&\qw &\qw & \qw&
\ghost{ {C}_2}}
  \end{aligned}
\\
\!\!\!\! \underbrace{ \qquad \qquad \qquad \qquad \qquad \qquad \qquad
  \quad} \\
\!\!\!\! S ^{(2)}
  \end{array}.
\nonumber
\end{align}
We adopt the convention that if wire
$i \in \mathcal{R}^{(N)}$ is connected with wire
$j \in \mathcal{S}^{(M)}$ they are identified by the same label,
i.e. $i=j$.
Again, if we want that such a composition
 forms a valid quantum strategy $\mathcal{R}^{(L)}_3$
we need to require that the graph of the connections in the composite
strategy does not contain closed loops.
By applying Proposition \ref{th:linkproduct}
it is possible to prove that
the comb of the composite network is given by the link product of 
$R ^{(N)}$ and $S ^{(M)}$, i.e. $T ^{(L)} = R ^{(N)} * S ^{(M)}$.

Consider now the problem of 
discriminating between two deterministic quantum strategy
$\mathcal{R}^{(N)}_0$ and $ \mathcal{R}^{(N)}_1$ given with prior
probability $\frac12$.
A possible way could be: i) prepare a multipartite input state,
possibly
entangled with some ancillary degrees of freedom, ii) send it
as input through the free input wires of the unknown strategy and eventually
iii) perform a two outcome POVM on the output state.
However, it is possible to exploit the causal order of the
quantum strategy so that the input at step $k$
can depend on the previous outputs at steps $j < k$, i.e
\begin{align}
\begin{aligned}
\Qcircuit @C=1em @R=1em {
\multiprepareC{1}{\rho}&\qw&\multigate{1}{D}&\qw&\multimeasureD{1}{\vec{P}}\\
 \pureghost{\rho}&\multigate{1}{C_1}&\ghost{D}&\multigate{1}{C_2}& \ghost{\vec{P}}\\
&\pureghost{C_1}&\qw&  \ghost{C_2} &
}    
  \end{aligned}\quad.
\end{align}
The most general way
for the discrimination of two deterministic quantum strategy
$\mathcal{R}^{(N)}_0$ and $ \mathcal{R}^{(N)}_1$
is then described 
by a two outcome tester
$\vec{\mathcal{T}^{(N+1)}} = \{ \mathcal{T}^{(N+1)}_0, \mathcal{T}^{(N+1)}_1 \}$
and the probability  of error $p_e$ 
as a function of $\mathcal{R}^{(N)}_0$, $ \mathcal{R}^{(N)}_1$ and $\vec{\mathcal{T}^{(N+1)}}$
is given by
\begin{align}
  \label{eq:errordiscrim}
  \begin{aligned}
 & p_e(\mathcal{R}^{(N)}_1, \mathcal{R}^{(N)}_0,
 \vec{\mathcal{T}^{(N+1)}}) 
= \\
& \frac12
(R^{(N)}_1* T ^{(N+1)}_0 +R_0^{(N)}* T ^{(N+1)}_1).  
  \end{aligned}
\end{align}
This leads to an operational notion of distance between quantum
strategies \cite{memeff}.
\begin{definition}
Let $\mathcal{R}^{(N)}_0$ and   $\mathcal{R}^{(N)}_1$
be two deterministic quantum strategies. The distance between
$\mathcal{R}^{(N)}_0$ and   $\mathcal{R}^{(N)}_1$ is given by
\begin{align}
  \label{eq:opnorm}
  \begin{aligned}
&  \| \mathcal{R}^{(N)}_0 - \mathcal{R}^{(N)}_1 \|_{op} := \\
&1 -2 \max_{\vec{\mathcal{T}^{(N+1)}}}
   p_e(\mathcal{R}^{(N)}_1, \mathcal{R}^{(N)}_0, \vec{\mathcal{T}^{(N+1)}})  
  \end{aligned}
\end{align}
where $\vec{\mathcal{T}^{(N+1)} } = \{ \mathcal{T}^{(N+1)}_0, \mathcal{T}^{(N+1)}_1\}$ is a tester
and $p_e$ is defined according to Eq. \eqref{eq:errordiscrim}.
\end{definition}
It is easy to prove that 
when 
$\mathcal{R}^{(N)}_0$ and   $\mathcal{R}^{(N)}_1$ are channels,
Eq. \eqref{eq:opnorm} leads to
the usual cb-norm distance.

\section{Memory cost of quantum strategies}\label{sec:entcost}
The main achievement of the
 general theory of quantum combs
is that arbitrarily complex quantum strategies
can always be represented by positive operators
subjected to linear constraints.
 This result is extremely relevant for
applications. Suppose we fix an information-processing task
and we look for the
quantum strategy that achieves the best performances
allowed by quantum theory.
Thanks to Theorem \ref{th:cornerstonetheorem}
this search is reduced to an optimization problem over
 a (convex) set  of suitably normalized positive operators.
Such a procedure is much more efficient than
separately optimizing each element of the strategy.

However, once the optimal Choi operator
has been found, one has to find an actual realization
of the quantum strategy.
Since a single quantum strategy
can be realized into many different ways one could be interested in finding
 the one that best fits some requirements.
For example, a reasonable request is to minimize the usage of some resource,
like the number of C-not gates.
Another resource which is valuable and hard to realize
in present day quantum technology is quantum memory.
One would benefit a lot from knowing how much quantum memory 
is needed in order to realize a given quantum strategy 
and whether it is possible to replace some quantum memory
with classical memory. 

In this section we provide an algebraic 
characterization of the amount of quantum memory which
is employed in the realization of a given quantum strategy.
As we pointed out in the previous section,
if $\boldsymbol{\mathscr{C}}$ is a realization of a deterministic
quantum strategy
$\mathcal{R}^{(N)}$,
the amount of memory which one has to 
preserve from step $i$ to step $i+1$
can be quantified by the dimension of 
the Hilbert space $\hilb{A}_i$.
Since we are interested in quantifying the amount of quantum memory
we need to introduce
a formalism that enables a distinction between quantum
memory and classical memory.
To this end, it is convenient to model a classical memory
as quantum system whose states must stay
diagonal with respect to
a fixed orthonormal basis $\{ \ket{i} \}$.
We then suppose that each  $\hilb{A}_i$ 
is split as
$\hilb{A}_i := \hilb{A}_i^{(c)} \otimes \hilb{A}_i^{(q)}$
where $\hilb{A}_i^{(q)}$ is the quantum memory and
$\hilb{A}_i^{(c)}$ is the Hilbert space that can carry only classical
information \cite{notedirectsum}.
With this definition Eq. \eqref{Eq:comb} becomes
\begin{align}
&
R^{(N)} = \begin{aligned}
    \Qcircuit @C=1em @R=1.5em {
\ustick{0}
&\multigate{2}{C_1} 
&\ustick{1} \qw 
&\ustick{2}
&\multigate{2}{C_{2}} 
&\ustick{3} \qw
&&&
&\ustick{2N-2}
&\multigate{2}{C_{N}}
& \ustick{2N-1} \qw
\\
&\pureghost{C_1}
&\ustick{\mathcal{A}_1^{q}}\qw
& \qw 
&\ghost{C_{2}}
& \ustick{\mathcal{A}_2^{q}} \qw
&&\dots&
&\ustick{\! \! \!  \! \mathcal{A}_{N-1}^{q}}
&\ghost{C_{N}}
&
\\
&\pureghost{C_1}
&\ustick{\mathcal{A}_1^{c}}\cw 
& \cw 
& \pureghost{C_{2}} \cw
&  \ustick{\mathcal{A}_2^{c}}\cw
&&\dots&
& \ustick{\!\!\! \!\mathcal{A}_{N-1}^{c}}
& \pureghost{C_{N}} \cw
&
}
  \end{aligned} ,
\end{align}
where the classical memories are denoted by double wires.

For the purpose if introducing the next two definitions, let
$\mathcal{R}^{(N)}$
be a deterministic Quantum Network, $\defset{S} = \{1,\dots , N\}$ be
its set of steps and $\defset{J}$
be a subset of $\defset{S}$.  We say that $\mathcal{R}^{(N)}$ can be
realized with $\vec{d} := \{ d_k \} $-dimensional quantum memories at
steps $\defset{J}$ if and only if there exists a realization
$\boldsymbol{\mathscr{C}}$ of $\mathcal{R}^{(N)}$ such that
$\dim(\hilb{A}_k^{(q)}) \leq d_k$ for all $k \in \defset{J}$.

\begin{definition} \label{def:zerr-quantum-cost} The \emph{zero error
memory cost at steps $\defset{J}$} of a deterministic
quantum strategy $\mathcal{R}^{(N)}$ is defined as
\begin{align}\label{eq:zerocost}
  \mathsf{M}_\defset{J} (\mathcal{R}^{(N)}, 0) := \min_{\boldsymbol {\mathscr{C}}} \max_{k \in \defset{J}}
 \log( \dim(\mathcal{A}^q_{k}) ) 
\end{align}
where the minimum is taken over all the possible
realization $\boldsymbol{\mathscr{C}}$ of $\mathcal{R}^{(N)}$.
\end{definition}

For any $\epsilon \geq 0$ it is possible to introduce the following notion.

\begin{definition} \label{def:eps-quantum-cost} The
  \emph{$\epsilon$-tolerant memory cost at steps $\defset{J}$}
  of $\mathcal{R}^{(N)}$ is defined as
\begin{align}
  \label{eq:epsiloncost}
 \mathsf{M}_{\defset{J}} (\mathcal{R}^{(N)}, \epsilon) := 
\min_{ \mathcal{S}^{(N)}\in   \mathrm{B}_{op}(\mathcal{R}^{(N)}, \epsilon) }
 \mathsf{M}_{\defset I} (\mathcal{S}^{(N)},0)
\end{align}
where $\mathrm{B}_{op}(\mathcal{R}^{(N)}, \epsilon)$
is the set of quantum strategies that are $\epsilon$-close to
$\mathcal{R}^{(N)}$ in the operational norm, i.e.
\begin{align}
&  \mathrm{B}_{op}(\mathcal{R}^{(N)}, \epsilon) :=
\{
\mathcal{S}^{(N)} \mbox{ s.t } \| \mathcal{S}^{(N)}-\mathcal{R}^{(N)} \|_{op} \leq \epsilon
\} \nonumber
\end{align}
where $\mathcal{S}^{(N)}$ is a deterministic quantum strategy.
\end{definition}

Eq. \eqref{eq:zerocost} quantifies the minimum amount of
 quantum memory  that one needs in order to realize
a given a quantum strategy $\mathcal{R}^{(N)}$.
In the case of a two steps deterministic
quantum strategy whose 
entanglement cost is zero 
we recover the notion of
 one-way  Local Operations and Classical Communication (LOCC).
More generally one could wonder how much quantum memory
is needed in the realization of a strategy
$\mathcal{S}^{(N)}$ which is similar to a target
one $\mathcal{R}^{(N)}$:
this intuition is formalized by Eq. \eqref{eq:epsiloncost}.

The following result \cite{algorithm} provides the least upper bound
to the amount of quantum memory which is required in the realization
of any deterministic quantum strategy where coherence is
preserved until the last step.
 \begin{proposition}
Any deterministic quantum strategy $\mathcal{R}^{N}$
can be realized with $\vec{d} := \{ d_k \} $-dimensional quantum memories at
steps $\defset{S}$, where
$d_k = \rank({R^{(k)}}) $.
 \end{proposition}

The main result of this section is
a necessary and sufficient condition for
a deterministic quantum strategy to be realized
with $\vec{d} := \{ d_k \} $-dimensional quantum memories at
steps $\defset{J}$. 
We first consider  the case  in which the set
$\defset{J}=\{k\}$ contains just a single step $k$, and then
we generalize the result to arbitrary sets.
Let us start with the following technical definition.

\begin{definition}
A quantum strategy
   $\mathcal{Q}^{(N)} \in
\mathcal{L}
(
\mathcal{L}(\hilb{H}_{\rm in}), 
\mathcal{L}(\hilb{H}_{\rm out})
)
$ is   \emph{deterministic  after the
  $k$-th step} if ${Q}^{(N)}$ satisfies
\begin{align}\label{detafterkstep}
& \Tr_{2l-1} [ Q^{(l)}] = I_{2l-2} \otimes Q^{(l-1)} \qquad l=k+1,
\dots, N~ \nonumber \\
& Q^{(k)} \leq R
\end{align}
where $R \in \mathcal{L}(\bigotimes_{i=0}^{2k-1}\hilb{H}_i) $ is a
deterministic quantum comb. 
\end{definition}
We are now ready to prove the following Proposition.
\begin{proposition} \label{th:quantum-cost}
A deterministic quantum strategy 
$\mathcal{R}^{(N)} \in
\mathcal{L}
(
\mathcal{L}(\hilb{H}_{\rm in}), 
\mathcal{L}(\hilb{H}_{\rm out})
)
$, can be
realized with a $d$-dimensional quantum memory at
step $k$ if and only if there exists a set $\{ \mathcal{Q}^{(N)}_j \}$
of quantum strategies deterministic after the $k$-th step such that
$R^{(N)} = \sum_j Q^{(N)}_j$ and $\rank(Q_j^{(k)}) \leq d$
\end{proposition}

\Proof
First we suppose that $R^{(N)}$ is
realizable with a $d$-dimensional quantum memory at
step $k$. Then there exists a set of channels
$\{ \mathcal{C}_i | \mathcal{C}_i :
  \mathcal{L}(\hilb{H}_{2i-2} \otimes \hilb{A}_{i-1}) \to
  \mathcal{L}(\hilb{H}_{2i-1} \otimes \hilb{A}_{i}) \}$ 
such that
$    C_1 * \cdots*C_{k}*C_{k+1}*\cdots * C_N = R^{(N)}$
and
$\hilb{A}_k := \hilb{A}_k^{(q)} \otimes \hilb{A}_k^{(c)}$
with $\dim(\hilb{A}_k^{(q)})=d$.
If we introduce the notation
$S:= C_1 * \cdots*C_{k}$ ($S \in \mathcal{L}(\bigotimes_{i=0}^{2k-1}
\hilb{H}_i \otimes \hilb{A}_k )$),
$T:= C_{k+1} * \cdots*C_{N}$ ($T \in \mathcal{L}( \bigotimes_{i=2k}^{N}
\hilb{H}_i \otimes \hilb{A}_k)$) we have
$    S*T = R^{(N)}$.
Let now $\mathcal{D} : \mathcal{L}(\hilb{A}_{k}^{(c)} ) \to
  \mathcal{L}( \hilb{A}_{k}^{(c)} )$ be 
the measure-and-reprepare channel on the classical system, whose  Choi operator is
$D
= \sum_i \ketbra{i}{i} \otimes \ketbra{i}{i}$.
Since the classical information is not affected by the action of 
$\mathcal{D}$ we have
\begin{align}
&
R = \sum_i \quad \begin{aligned}
    \Qcircuit @C=1.2em @R=1em {
       &&\multigate{2}{C_k} &\qw
\qw&
&\multigate{2}{C_{k+1}} & \qw&
\\
   \dots&   &\pureghost{C_k} \qw &
\qw & \qw 
      &\ghost{C_{k+1}}&  \qw
&
\dots
\\
\dots &&\pureghost{C_k} \cw &
\puremeasureD{i} \cw &\prepareC{i} \cw & 
      \pureghost{C_{k+1}} \cw&  \cw
&
\dots
}
  \end{aligned} 
\nonumber \\
\nonumber\\
&\begin{aligned}
  & R^{(N)} =  S*T = S*D*T = \\
&S*\sum_i \ketbra{i}{i} \otimes \ketbra{i}{i}* T = \sum_i S_i * T_i
\label{eq:defSiandTi}
\end{aligned}
\end{align}
where $S_i = S * \ketbra{i}{i}$ and $T_i = T * \ketbra{i}{i}$.
 We have that the set
 $\{ S_i \}$ defines a generalized instrument
while $T_i$ defines a deterministic quantum strategy for each $i$. 
Let us now  consider
the spectral decompositions
of the operators $S_i$,
\begin{align}
 S_i  = \sum_{j\in J_i} X_{j,i}  
\qquad
X_{j,i} :=
\ket{\psi_{j,i}}\bra{\psi_{j,i}}
\label{eq:specdecompSi}
\end{align}
where $J_i$ are disjoint sets.  Notice that the set $\{ X_{j,i} \}$
defines a generalized instrument from which $\{ S_{i} \}$ can be
obtained by coarse graining.  Let us now define $Q^{(N)}_{j,i} :=
X_{j,i}*T_{i}$.  One can verify that $Q^{(N)}_{j,i}$ is deterministic
after the $k$-th step for all $j,i$. Since
$Q^{(k)}_{j,i}=\Tr_{A^{(q)}_k}(X_{j,i})=\Tr_{A^{(q)}_k}(\ketbra{\psi_{j,i}}{\psi_{j,i}})$
the dimension of $\hilb{A}^{(q)}_k$ is an upper bound for the Schmidt
rank of $\ket{\psi_{j,i}}$ with respect the bipartition
$(\bigotimes_{i=0}^{2k-1}\hilb{H}_i) \otimes \hilb{A}^{(q)}_k$, which
consequently implies that the rank of $Q^{(k)}_{j,i}$ is at most $d$.
Combining Eqs. \eqref{eq:defSiandTi} and \eqref{eq:specdecompSi} we
have $\sum_{ij} Q^{(N)}_{i,j} = \sum_i(\sum_j X_{j,i})*T_i = \sum_i
S_i * T_i = R^{(N)}$ and the thesis is proved.

We now prove the sufficiency of the condition.
By hypothesis we have $R^{(N)} = \sum_j Q^{(N)}_j $ where
 the $\{ Q^{(N)}_j \}$ are deterministic after the $k$-th step.
Let us introduce the operators
$ \KetBra{Q^{(k)\frac12}_j}{Q^{(k)\frac12}_j} \otimes \ketbra{j}{j} \in
\mathcal{L}(\bigotimes_{i=0}^{2k-1}\hilb{H}_i \otimes \mathcal{A}^{(q)}_{k,j}
\otimes \mathcal{A}^{(c)}_k )$
where $\mathcal{A}^{(q)}_{k,j} := \supp(Q^{(k)}_j)$
and $\mathcal{A}^{(c)}_k$ is an Hilbert space carrying classical
information encoded into the orthonormal basis
$\ket{j}$.
Since $\rank(Q^{(k)}_j) \leq d$ for each $j$ we can without loss of generality
consider
an isometric embedding of  each
$\mathcal{A}^{(q)}_{k,j}$ into a $d$ dimensional
Hilbert space $\mathcal{A}^{(q)}_{k}$.
One can easily check that
$S := \sum_j\KetBra{Q^{(k)\frac12}_j}{Q^{(k)\frac12}_j} \otimes \ketbra{j}{j}$
satisfies the normalization
\eqref{recnorm}
 and then  Theorem
\ref{th:cornerstonetheorem}
implies that there exists a realization
$S = C_{1} * \dots * C_{k} $
where
 $C_{k} \in \mathcal{L}(\mathcal{A}_{k-1}
 \otimes
\mathcal{H}_{2k-2} \otimes 
\mathcal{A}_{k}
 \otimes
\mathcal{H}_{2k-1})$.

We now introduce the operator
$T := \sum_j\ketbra{j}{j} \otimes Q_j^{(k)-\frac12} Q_j^{(N)} Q_j^{(k)-\frac12}
\in \mathcal{L}(
\mathcal{A}^{(c)}_k
 \otimes
\mathcal{A}^{(q)}_{k}
\otimes
\bigotimes_{i=2k}^{2N-1}\hilb{H}_i ) $
(also in this case we assumed the embedding
$\mathcal{A}^{(q)}_{k,j} \hookrightarrow \mathcal{A}^{(q)}_{k}$).
One can prove that 
$T$ is a well defined 
deterministic quantum comb.
There exists then
a realization $T = C_{k+1} * \dots * C_N$
where
 $C_{k+1} \in \mathcal{L}(\mathcal{A}_{k}
 \otimes
\mathcal{H}_{2k} \otimes 
\mathcal{A}_{k+1}
 \otimes
\mathcal{H}_{2k+1})$.
It is easy to verify that $S*T = R^{(N)}$
which in turns implies
that $C_1 * \dots * C_k * C_{k+1} * \dots * C_N $
is a realization of $R^{(N)}$ with $\dim{\mathcal{A}^{(q)}_k} = d$.
\qed

The result of Proposition \ref{th:quantum-cost} can be extended to the
case of multiple steps.
 \begin{theorem}\label{th:maintheorem}
 Let
 $\mathcal{R}^{(N)}$
 be a deterministic quantum strategy
 and let $\defset{J} $
 be a set of steps.
For each $k \in \defset{J} $ 
we introduce an
index ${i}_{k}$.
 The following two statements are equivalent:
 \begin{itemize}
 \item $\mathcal{R}^{(N)}$ is 
 realizable with  $\vec{d} := \{ d_k \} $-dimensional quantum memories at
 steps $\defset{J}$.
 \item 
 there exists a set 
$Q^{(N)}_{\vec{i}}$, $\vec{i} = i_{k_{\rm min}},\dots, i_{k_{\rm max}}$  such that
 \begin{align}
   \begin{aligned}
&     R^{(N)} = \sum_{\vec{i}} Q^{(N)}_{\vec{i}},  
\qquad
\rank(Q_{i_{k_{ \rm min}}, \dots , i_{k} }^{(k)}) \leq d_k\\
& Q^{(N)}_{i_{k_{ \rm min}}, \dots , i_{k} } \mbox{ are deterministic after
  the }
k 
\mbox{ step,} 
   \end{aligned} \nonumber
 \end{align}
where we defined
\begin{align}
Q^{(N)}_{i_{k_{ \rm min}}, \dots ,i_{k} } 
:= \sum_{i_{k'}} Q^{(N)}_{i_{k_{ \rm min}},\dots , i_{k'} }.\nonumber  
\end{align}
with $k'$ denoting the element following $k$ in $\defset{J}$.
 \end{itemize}
\end{theorem}
\Proof
The result follows by iterating the proof of Proposition \ref{th:quantum-cost}.
\qed

One could wonder whether 
the existence of a realization of a quantum strategy
$\mathcal{R}^{(N)}$ with memory $d_k$ at step $k$
and a of realization
with memory 
$d_l$ at step $l$,
 implies that there exists a realization of
$\mathcal{R}^{(N)}$ with
$\{ d_k , d_l \} $-dimensional quantum memories at
steps $\{k, l\}$.
This would imply the equality 
$
 \mathsf{M}_{ \defset{J}\cup\defset{I} } (\mathcal{R}^{(N)}, 0) =    \max\{\mathsf{M}_{ \defset{J} }(\mathcal{R}^{(N)}, 0), \mathsf{M}_{ \defset{I}  } (\mathcal{R}^{(N)}, 0)\}  
$
for any two disjoint sets of steps $\defset{J} ,\defset{I} \subseteq \defset{S}$.
If this were true, a global minimization of the quantum memory would reduce
to $N-1$ independent minimizations at each step.
Unfortunately this is not the case, as shown by the following
counterexample.

Bennett et al. in Ref. \cite{unextprodbas}
  introduced 
a state $\rho \in \mathcal{L}(\hilb{H}_0 \otimes \hilb{H}_1 \otimes \hilb{H}_2  )$
which is two-way separable but not three-way
separable, i.e. we have
$\rho = \sum_i \sigma_i^{[01]} \otimes \tau_i^{[2]} = \sum_j \tilde{\rho}_j^{[0]} \otimes \tilde{\tau}_j^{[12]}$
for some unnormalized states
but we cannot have
$\rho = \sum_i \alpha_i^{[0]} \otimes \beta_i^{[1]} \otimes
\gamma_i^{[2]} $
for some others
unnormalized states
\cite{notetristate}.
Every normalized quantum state can be 
interpreted as quantum strategy with trivial input
spaces, and thus we have 
\begin{align}
&\rho = 
\begin{aligned}
\Qcircuit @C=0.4em @R=0.05em {
    &\multipuregate{1}{\;}&\qw&&\multipuregate{1}{\;}&\qw &&\multipuregate{1}{\;}&\qw\\
    &\pureghost{\;}&\cw&\cw&\pureghost{\;}\cw&\qw&\qw&\ghost{\;}&&}        
\end{aligned}
\mbox{ or }\ \rho = 
\begin{aligned}
\Qcircuit @C=0.4em @R=0.05em {
    &\multipuregate{1}{\;}&\qw&&\multipuregate{1}{\;}&\qw &&\multipuregate{1}{\;}&\qw\\
    &\pureghost{\;}&\qw&\qw&\ghost{\;}&\cw&\cw&\pureghost{\;}\cw&&}  
\end{aligned}, 
  \nonumber \\
&\mbox{but } \rho \neq 
\begin{aligned}
\Qcircuit @C=0.4em @R=0.05em {
    &\multipuregate{1}{\;}&\qw&&\multipuregate{1}{\;}&\qw &&\multipuregate{1}{\;}&\qw\\
    &\pureghost{\;}&\cw&\cw&\pureghost{\;}\cw&\cw&\cw&\pureghost{\;}\cw&&}      
\end{aligned}
\nonumber \quad.
\end{align}
The fact that $\rho$
is two-way separable but not three-way separable
means that the three steps quantum strategy $\rho$
is realizable with $1$-dimensional quantum memory
either at step $1$ or at step $2$
but it cannot be realized with 
$1$-dimensional quantum memory
at both steps, i.e.
\begin{align}
  \label{eq:incompatibility} 
 \mathsf{M}_{ \{ 1, 2 \} } (\rho, 0) >   \max\{ \mathsf{M}_{ \{ 1 \} }
 (\rho, 0) ,  \mathsf{M}_{ \{ 2 \} } (\rho, 0) \}.
\end{align}
Moreover, we notice that 
it is possible to build a whole class of $3$-step
quantum strategies with the property
\eqref{eq:incompatibility}
by linking an isometric channel  to each subsystem of $\rho$, i.e.
 \begin{eqnarray}
&\mathcal{S}^{(3)} = 
  \begin{array}{c}
\begin{aligned}
\Qcircuit @C=0.4em @R=0.05em {
&&\ghost{V_1}&\qw&& \ghost{V_2}&\qw&&\ghost{V_3}&\qw\\
    &\multipuregate{1}{\;}&\multigate{-1}{V_1}&&\multipuregate{1}{\;}&\multigate{-1}{V_2} &&\multipuregate{1}{\;}&\multigate{-1}{V_3}\\
    &\pureghost{\;}&\qw&\qw&\ghost{\;}&\cw&\cw&\pureghost{\;}\cw&&}  
\end{aligned}\\
\! \! \! \! \! \! \! \! \! \! \! \! \!  \underbrace{\qquad \qquad \qquad \qquad \qquad \quad}\\
\! \! \! \! \! \! \! \! \! \! \! \! \!  \rho       
  \end{array} \nonumber &\\
 &\mathsf{M}_{ \{ 1, 2 \} } (\mathcal{S}^{(3)}, 0) = 1, \qquad
    \mathsf{M}_{ \{ 1 \} }
 (\mathcal{S}^{(3)}, 0) =  \mathsf{M}_{ \{ 2 \} } (\mathcal{S}^{(3)},
      0) = 0. &\nonumber
\end{eqnarray}

\section{Examples and applications}\label{sec:examples}
It is in general a hard task to verify whether a
deterministic quantum strategy
 can be realized with a given amount of quantum memory
and to calculate its memory cost.
Nevertheless, some properties of the quantum comb
may imply non trivial bounds on the quantum memory
which is needed in the realization.

\subsection{Memory requirements in the presence of symmetry}

In this section we 
show that
if a quantum strategy 
enjoys some symmetries, then
the amount of quantum memory
needed in the realization  
can be efficiently bounded.
The following Proposition
provides the main tool we will use to prove such a bound.

\begin{proposition}\label{pr:blockdiagandmemory}
Let
$
\mathcal{R}^{(N)} \in
\mathcal{L}
(
\mathcal{L}(\hilb{H}_{\rm in}), 
\mathcal{L}(\hilb{H}_{\rm out})
)
$
be a deterministic quantum strategy
and
$\{ P_i , P_i \in \mathcal{L}(\bigotimes_{i=0}^{2k-1}\hilb{H}_{i})\}$
be a set of orthogonal projectors
such that 
$\sum_i P_i = I_{0 \dots 2k-1}$
where $I_{0 \dots 2k-1}$ is the identity
on $\bigotimes_{i=0}^{2k-1}\hilb{H}_{i}$.
Suppose that $R^{(N)} = \sum_i P_i R^{(N)} P_i$.
Then $ \mathcal{R}^{(N)}$
is realizable with $d_k$-dimensional memory 
at step $k$
where $d_k := \max_i \Tr[P_i]$.
Moreover if $
\mathcal{R}^{(N)}$
is realizable with 
$d_l$-dimensional memory at step $l$
with $l > k$,
then 
$\mathcal{R}^{(N)}$
is also realizable
with $\{ d_k, d_l \}$-dimensional
memories at steps $\{k, l \}$ 
\end{proposition}

\Proof
Let us define
$Q_i^{(N)} := P_i R^{(N)} P_i$.
They satisfy
the hypothesis of Proposition \ref{th:quantum-cost}
with $\rank(Q_i^{(k)}) \leq d_k$.

Consider now the case in which
$\mathcal{R}^{(N)}$
is realizable with 
$d_l$ memory at step $l > k$.
Then there exists a set of operators
$\tilde{Q}^{(N)}_j$
satisfying the hypothesis of 
of Proposition \ref{th:quantum-cost}
with $\rank(\tilde{Q}_j^{(l)}) \leq d_l$.
Let us now define
$Q^{(N)}_{i,j} := P_i \tilde{Q}_j^{(l)} P_i$.
One can verify
that they satisfy the hypothesis
of Theorem \ref{th:maintheorem}
with 
$\rank(Q^{(N)}_i) \leq d_k$ (we remind that
$Q^{(N)}_i := \sum_jQ^{(N)}_{i,j}$).
\qed

Before considering the case of 
quantum strategies
with symmetries let us now
 introduce some preliminary notions of group representation theory.
If  $U(g) \in \mathcal{L}(\hilb{H})$
is a unitary representation of a compact Lie group
then it is decomposable 
into a direct sum of irreducible representations
$U(g) = \bigoplus_\nu U_\nu(g) \otimes I_{m_\nu}$,
where $U_\nu(g) \in \mathcal{L}(\hilb{H}_\nu)$
and $\hilb{H} = \bigoplus_\nu \hilb{H}_\nu \otimes \mathbb{C}^{m_\nu}$.
The spaces $\hilb{H}_\nu$'s are customarily called representation
spaces while the $\mathbb{C}^{m_\nu}$'s
are called multiplicity spaces. We are now ready to prove the main result of this section.
\begin{proposition}
\label{th:symindcom}
Let $\mathcal{R}^{(N)} \in \mathcal{L}(\bigotimes_{i=0}^{N}\hilb{H}_i)$ be 
a deterministic quantum strategy and let $U(g) \in
\mathcal{L}(\bigotimes_{i=0}^{2k-1}\hilb{H}_i)$,
$U(g) = \bigoplus_\nu U_\nu(g) \otimes I_{m_\nu}$,
be a unitary representation of a compact Lie group $G$.
If the commutation
\begin{align}
[R^{(N)}, I_{2N-1\ldots 2k}\otimes U(g) ]=0 \quad \forall g\in G
\label{eq:defsymmetry}
\end{align}
holds
then $\mathcal{R}^{(N)}$ is realizable with $d_k$ dimensional
quantum memory at step $k$ where
$d_k$ is 
the dimension of the largest multiplicity space, i.e
$d_k := \max_\nu m_\nu$
\end{proposition}
\Proof
Eq. \eqref{eq:defsymmetry} and
the Schur's lemmas imply
the decompositions
\begin{align}\label{eq:decompochoi}
R^{(N)}  = \sum_\nu P_\nu \otimes r_\nu
\end{align}
Let $\{ \ket{\psi_\nu^{j}} \}$ be an orthonormal basis
for $\hilb{H}_\nu$ and let $P_{m_\nu}$
denote the projectors on the multiplicity
spaces
$\mathbb{C}^{m_\nu}$.
We now define the projectors
$P_{\nu,j} := 
\ketbra{\psi_\nu^{j}}{\psi_\nu^{j}} \otimes P_{m_\nu}$.
Since we have $\sum_{\nu,j }P_{\nu,j} = I_{0 \dots 2k-1}$
and Eq. \eqref{eq:decompochoi} implies
$R^{(N)} := \sum_{\nu,j}  P_{\nu,j} R^{(N)} P_{\nu,j}$,
the conditions of Proposition
\ref{pr:blockdiagandmemory}
are satisfied with
$d_k := \max_{\nu,j} \Tr[P_{\nu,j}] = \max_{\nu} m_\nu$.
\qed

The optimal
cloning of a unitary transformation for any dimension $d \geq 2$ \cite{cloning}
provides an example of a quantum strategy $\mathcal{R}^{(2)}$ 
that enjoys the property \eqref{eq:defsymmetry},
with $\max_\nu m_\nu = 2$.
We therefore conclude that any covariant protocol
for cloning unitary operators has a memory cost of one qubit,
independently on the dimension.

\subsection{Memory cost of quantum channels}

The aim of this section 
is to 
specialize the notion of memory cost to the case
of channels and to provide
examples that
allow for an easy calculation.
Reminding Eq. \eqref{eq:channelasnetw}
a quantum channel $\mathcal{C}:\Lin(\Lin(\hilb{H}_{\rm in }) , \Lin
(\hilb{H}_{\rm out}))$
can be represented as a two-step deterministic quantum comb.
This interpretation
corresponds to decompose $\mathcal{C}$
 into  an encoding 
channel $\mathcal{C}_1:\Lin(\Lin(\hilb{H}_{\rm in }) , \Lin
(\hilb{A}^{q} \otimes \hilb{A}^{c} ))$
followed by a  decoding channel
$\mathcal{C}_2:\Lin(\Lin
(\hilb{A}^{q} \otimes \hilb{A}^{c}),
\Lin(\hilb{H}_{\rm out})  )$,
\begin{align}
&
\begin{aligned}
    \Qcircuit @C=1em @R=1em {
\ustick{{\rm in}}&\gate{C}&\ustick{{\rm out}}\qw& }
  \end{aligned}
\quad
 =
 \quad
\begin{aligned}
    \Qcircuit @C=1em @R=1em {
\ustick{{\rm in}}
&\multigate{2}{C_1} 
&
&
&\pureghost{C_{2}} 
&\ustick{{\rm out}} \qw
\\
&\pureghost{C_1}
&\ustick{\mathcal{A}^{q}}\qw
& \qw 
&\ghost{C_{2}}
& 
\\
&\pureghost{C_1}
&\ustick{\mathcal{A}^{c}} \cw 
& \cw 
& \multipuregate{-2}{C_{2}} \cw
& 
}
  \end{aligned} .
\end{align}
Applying Definition \ref{def:zerr-quantum-cost},
we say that a quantum channel
$\mathcal{C}$
is realizable with 
$d$-dimensional quantum memory 
when there exist
an encoding channel
channel $\mathcal{C}_1:\Lin(\Lin(\hilb{H}_{\rm in }) , \Lin
(\hilb{A}^{q} \otimes \hilb{A}^{c} ))$
and a  decoding channel
$\mathcal{C}_2:\Lin(\Lin
(\hilb{A}^{q} \otimes \hilb{A}^{c}),
\Lin(\hilb{H}_{\rm out})  )$ 
such that $\dim(\mathcal{A}^q) \leq d$
and $C = C_1 * C_2$.
Thanks to
Proposition
\ref{th:quantum-cost},
this holds true if and only if
there exists a set of operators $\{Q_i \}$ 
such that $C = \sum_i Q_i$
and $\rank(\Tr_{\rm out}[Q_i]) \leq d$.
It is easy to verify that there is no loss of generality 
if we assume  $\rank(Q_i) = 1$.
We have then that
a quantum channel
$\mathcal{C}$
is realizable with 
$d$-dimensional quantum memory 
when there exist a decomposition
$C = \sum_K \KetBra{K}{K}$
such that $\rank(K^\dagger K) \leq d$.
The zero-error memory cost $\mathsf{M}(\mathcal{C}, 0)$
is equivalent to the zero-error entanglement cost
of the quantum state ${d_{\rm in}^{-1}} C$ \cite{buscnila}.

A similar notion of memory cost of quantum channel $\mathscr{E}(\mathcal{C})$ has been recently
introduced in Ref. \cite{zurich} and can be rephrased within our
framework as follows:
\begin{align}
  \mathscr{E}(\mathcal{C}) = 
\lim_{\substack{\epsilon \to 0 \\ n \to \infty}} \frac1n
\mathsf{M}(\mathcal{C}^{\otimes n}, \epsilon).
\end{align}
In Ref. \cite{zurich}
the authors proved that the quantity
$  \mathscr{E}(\mathcal{C})$
can be expressed in terms of the entanglement of formation
and they discuss the relation between 
$  \mathscr{E}(\mathcal{C})$ and the quantum channel capacity of
$\mathcal{C}$.

In the previous section we discussed the relation between 
symmetry properties and quantum memory. We now
 consider two particular classes of covariant channels
which allow for an easy calculation of the zero-error memory
cost.
This is the case of covariant channels $\mathcal{C} \in \Lin(\Lin(\hilb{H}),\Lin(\hilb{H}))$ satisfying
the constraints
\begin{align}
  &\mathcal{C}(U\rho U^\dag)=U\mathcal{C}(\rho)U^\dag,\label{eq:cov}\\
  &\mathcal{C}(U^*\rho U^T)=U\mathcal{C}(\rho)U^\dag,\label{eq:con}
\end{align}
 One can prove that 
condition \eqref{eq:cov}
implies the  following form for the Choi operator
\begin{align}
  C_\alpha:=&\alpha \frac1d|I\kk\bb I|+\beta \left(I-\frac1d|I\kk\bb I|\right)\label{eq:covch}
  \end{align}
where $\alpha+(d^2-1)\beta=d$. On the other hand
Eq. \eqref{eq:con} implies
  \begin{align}
  C_\gamma :=&\gamma P_+ +\delta
  P_-,\label{eq:conch}
\end{align}
where $P_\pm=1/2(I\pm E)$ are the projections on the symmetric and
anti-symmetric space of $\hilb{H} \otimes \hilb{H} $, respectively, $E$ is the swap operator
$E|\varphi\>|\psi\>=|\psi\>|\varphi\>$ and
$(d+1)\gamma+(d-1)\delta = 2$.

In the case of a symmetry as
in Eq.~\eqref{eq:covch}, the zero error entanglement cost of states
$1/d C_\alpha$ was evaluated in Ref. \cite{horoter}.
This result implies that $\mathsf{M}(C_\alpha,0) = \log(\lceil \alpha
\rceil )$
where $\lceil \alpha
\rceil$  denotes the ceiling of $\alpha$.

As regards the case of Eq.~\eqref{eq:conch}, one realizes
that $C_\gamma$ are rescaled Werner states \cite{wernerstate} by a factor $d$.
Thus for $1/(d+1) \leq \gamma \leq 2/(d+1)$
$C_\gamma$ is a seperable operator and consequently
$\mathsf{M}(C_\gamma,0) = 0$.
Since $P_\pm$ can be decomposed as
the sum of rank one projections on the states $|m\>|m\>$ and
$1/\sqrt2(|m\>|n\>\pm|n\>|m\>)$, whose partial trace
$1/2(|m\>\<m|+|n\>\<n|)$ has rank $2$, we always have
$\mathsf{M}(\mathcal{C}
_\alpha,0)=1$,
when $0 \leq \gamma \leq 1/(d+1)$, irrespectively of the dimension $d$.

\section{Conclusions}\label{sec:concl}

 In conclusion, we defined the notion of memory cost for a
quantum strategies, that captures the minimal dimension of ancillary
systems that needs to be kept coherent during an algorithm specified
by the comb  representing the strategy. The realization of the
strategy using minimal global ancillary dimension can be algebraically
characterized by theorem \ref{th:maintheorem}, representing our main
result. 

 We also showed by an example that the optimization of the memory required
between two steps of the computation is in general not compatible with
the optimization of the memory required between two different steps.

We notice that the algebraic condition provided by Theorem \ref{th:maintheorem}
does not allow for an easy evaluation of the
memory cost for a given strategy.
Because of that, providing non-trivial bound on the memory requirement
becomes an issue. In this paper we showed that
symmetry arguments can help to calculate
the memory cost of some particular channels and strategies, like e.g.
the covariant cloning of unitary transformation.

A natural continuation of this line of research would be 
to look for other conditions that can provide similar bounds for the
memory cost.

\acknowledgments This work has been supported by EU FP7 programme
through the STREP project COQUIT
and
by the Italian Ministry of Education through PRIN 2008.


\begin{thebibliography}{99}
\bibitem{note1} If a device is used more than once we represent each
  different use with a linear map so that different uses of the same
  device are actually different linear maps.
\bibitem{note2} Within our formalism a physical closed loop in the
  lab, that is taking the output of a device and then sending it as an
  input to the same device, corresponds to many subsequent uses of the
  same transformation.
\bibitem{notedirectsum}
This formalism is equivalent to the usage of a direct sum over the
classical label $\mathcal{A} = \bigoplus_i \mathcal{A}^{(i)}$ with the
prescription that each Hilbert space
$\mathcal{A}^{(i)}$
 is embedded into an Hilbert space
$\tilde{\mathcal{A}}^{(i)}$ with
$\dim(\tilde{\mathcal{A}}^{(i)}) = \max_j \dim(\mathcal{A}^{(j)})$.
\bibitem{notetristate} The superscript $[ij]$ denotes that $\xi^{[ij]}
  \in \mathcal{L}(\hilb{H}_i \otimes \hilb{H}_j)$.
\end{thebibliography}
\end{document}